\documentstyle[12pt]{article}
\newcommand{\be}{\begin{equation}}
\newcommand{\ee}{\end{equation}}
\begin{document}
\def\theequation{\arabic{section}.\arabic{equation}}
\begin{titlepage}
\title{Inflation and quintessence with nonminimal coupling}
\author{Valerio Faraoni$^{1,2}$\\ \\
{\small \it $^1$~Research Group in General Relativity (RggR)} \\ 
{\small \it Universit\'e Libre de Bruxelles,  Campus Plaine CP 231} \\
{\small \it  Blvd. du Triomphe, 1050 Bruxelles, Belgium}\\ \\
{\small \it $^2$ INFN-Laboratori Nazionali di Frascati, 
P.O. Box 13, I-00044 Frascati, Roma (Italy)} 
}
\date{} 
\maketitle   \thispagestyle{empty}  \vspace*{1truecm}
\begin{abstract} 
The nonminimal coupling (NMC) of the scalar field to the Ricci curvature
is 
unavoidable in many cosmological scenarios. Inflation and quintessence
models based on 
nonminimally coupled scalar fields are studied, with particular
attention to the
balance between the scalar potential and the NMC term
$\xi R \phi^2 /2$ in the action. NMC makes
acceleration of the universe harder to achieve for the usual
potentials,  but it is beneficial in obtaining
cosmic acceleration with unusual potentials. 
The slow-roll approximation with NMC, conformal transformation
techniques, and other aspects of the
physics of NMC are clarified.
\end{abstract}
\vspace*{1truecm} 
\begin{center} 
\end{center}     
\end{titlepage}   \clearpage

\section{Introduction}

The idea of cosmological inflation is legitimately regarded as
a breakthrough of modern cosmology: it solves the horizon, flatness and
monopole problem, and it provides a mechanism for the generation of
density
perturbations needed to seed the formation of structures in the universe
\cite{KolbTurner,Lindebook}. The essential qualitative feature of
inflation,
the
acceleration of the universe, is also required (albeit at a different
rate) at the present epoch of the
universe in order to explain the data from high redshift supernovae
\cite{SN}. If confirmed, the latter imply that a form of matter with
negative
pressure (``quintessence'') is beginning to dominate the dynamics
of the universe. Scalar fields have been proposed as natural models of
quintessence \cite{Zlatevetal,Steinhardtetal,Chiba,Uzan,PBM}.

Inflation is believed to be driven by a scalar field, apart
possibly from the $R^2$ inflationary scenario in higher
derivative theories of gravity \cite{footnote1} or
in supergravity (e.g.  \cite{Starobinsky80,Maeda89}).

The inflaton  field $\phi$ obeys the 
Klein-Gordon equation \cite{footnote2}   
\setcounter{equation}{0}
\begin{equation}  \label{KG}
\Box \phi-\xi R \phi-\frac{dV}{d\phi}=0 \; , 
\end{equation}
where $V( \phi)$ is the scalar field potential, $R$
denotes the Ricci curvature of spacetime, and the term $-\xi R \phi$ in
Eq.~(\ref{KG}) describes the explicit nonminimal coupling (NMC) of
the field $\phi$ to the Ricci curvature \cite{CherTag,CCJ}. A possible
mass term $m^2 \phi^2
/2 $ for the
field $\phi$ and the cosmological
constant $\Lambda$ are embodied in the
expression of $V( \phi ) $.

Eq. (\ref{KG}) is derived from the Lagrangian density
\begin{equation}  \label{Lagrangiandensity}
{\cal L}\sqrt{-g} =\left[ \frac{R}{16\pi G}-\frac{1}{2} \nabla^{c}\phi  
\nabla_{c} \phi -V( \phi) - \frac{\xi}{2}R\phi^2 \right] \sqrt{-g}\; ,
\end{equation}
where  $g$ is the determinant of
the metric tensor $g_{ab}$, and $\nabla_{c}$ is the covariant
derivative 
operator. In inflationary theories it is assumed that the
scalar field dominates the evolution of the universe and that no forms of
matter other than $\phi$ are included in the Lagrangian density 
(\ref{Lagrangiandensity}).

Two values of the coupling constant $\xi$ are most often encountered in
the
literature: $\xi=0$ ({\em minimal coupling}) and $\xi=1/6$ ({\em
conformal coupling}) \cite{footnote3}, while the possibility
$\left| \xi
\right| >>1 $ ({\em strong coupling}) has also been considered many  
times, for both signs of $\xi$
\cite{MakinoSasaki91,FakirUnruh92ApJ,LTMI,FutamaseTanaka,
BassettLiberati,HwangNoh98,Chiba}.

Contrarily to common belief, the introduction of NMC is not
a matter of
taste; NMC is instead forced upon us in many situations of physical and
cosmological interest.  There are many compelling reasons to include an
explicit nonminimal (i.e. $\xi \neq 0$) coupling in the action: NMC 
arises at the quantum level when quantum corrections to the scalar
field theory are considered, even if $\xi=0$ for the classical,
unperturbed, theory \cite{BirrellDavies80}; NMC is necessary
for the renormalizability of the scalar field theory in curved space  
\cite{CCJ,FreedmanWeinberg74,FreedmanMuzinichWeinberg74}. But what is the
value of $\xi$ ? This
problem has been addressed at both 
the classical and quantum level (\cite{SonegoFaraoni93,Faraoni96, 
Buchbinderetal}, and 
references therein). The answer depends on the theory of gravity and of
the scalar field adopted; in most theories used to describe
inflationary scenarios, it turns out that a value of the coupling constant
$\xi \neq 0 $ cannot be avoided. 

In general relativity, and in all other metric theories of gravity in
which the scalar field $\phi$ is not part of the gravitational sector, the
coupling constant necessarily assumes the value $\xi=1/6$
\cite{SonegoFaraoni93,Gribetal,Faraoni96}. The study of  asymptotically
free theories in an external gravitational field, described by the
Lagrangian density
\be
{\cal L}_{AF} \sqrt{-g} =\sqrt{-g} \left( aR^2 +b \, G_{GB} +
c \, C_{abcd} C^{abcd} - \xi R
\phi^2 + 
{\cal L}_{matter} \right)
\ee
(where $G_{GB}$ is the Gauss-Bonnet invariant and $C_{abcd} $ is the
Weyl tensor) shows a scale-dependent 
coupling parameter $\xi ( \tau ) $. In Refs. \cite{Buchbinderetal,bookBOS}
it was shown that  asymptotically free grand unified theories (GUTs) have
a $\xi$ depending on a
renormalization group parameter $\tau$, and that $\xi ( \tau ) $
converges to $1/6$, $\infty$, or to any initial condition $\xi_0$  as
$\tau
\rightarrow \infty$ (this limit corresponds to strong curvature conditions
and to the early universe), depending on the gauge group and on the matter
content of the theory. In Ref. \cite{ParkerToms} it was also obtained
that $\left| \xi ( \tau ) \right| \rightarrow + \infty$ in $SU(5)$ GUTs.
Similar results were derived in finite GUTs without running of the gauge
coupling, with the convergence of $\xi$ to its asymptotic value being much
faster \cite{Buchbinderetal,bookBOS}. An exact renormalization group study
of
the $\lambda \phi^4 $ theory shows that $\xi=1/6$ is a stable infrared
fixed point \cite{Bonanno}.

In the large $N$ limit of the Nambu-Jona-Lasinio model, $\xi=1/6$
\cite{HillSalopek92}; in the $O(N)$-symmetric model with $V=\lambda
\phi^4$, $\xi$ is generally nonzero and depends on the coupling constants
$\xi_i$ of the individual bosonic components \cite{Reuter94}. Higgs fields
in the standard model have $\xi \leq 0$ or $\xi \geq 1/6$
\cite{Hosotani85}. Only a few investigations produce $\xi=0$: the minimal
coupling is predicted if $\phi$ is a  Goldstone boson with a spontaneously
broken global symmetry \cite{VoloshinDolgov82}, for a
semiclassical scalar field with backreaction and 
cubic self-interaction \cite{Hosotani85}, and for theories formulated in
the Einstein conformal frame \cite{Faraoni96,Calgary}. In view of the
above results, it is wise to incorporate
an explicit NMC between $\phi$ and $R$ in the inflationary paradigm and
in quintessence models.

A conservative approach to inflation and quintessence employs general
relativity as the
underlying gravity theory (exceptions are $R^2$, extended, hyperextended
and stringy inflation and the extended quintessence model of
Ref.~\cite{PBM}),
and conformal coupling 
is unavoidable in general relativity, as well as in any metric
theory of gravity in which the scalar field is part of  
the non-gravitational sector (e.g. when $\phi$ is a Higgs field) 
\cite{SonegoFaraoni93,Gribetal,Faraoni96}.

The viability of an inflationary 
scenario and the constraints on the inflationary model 
are profoundly affected by the presence of NMC and
by the value of the coupling constant $\xi$ 
(\cite{Faraoni96} and references therein;
\cite{FakirUnruh90a,FakirUnruh90b,FutamaseMaeda89}). The
analysis 
of the various inflationary scenarios considered in the literature usually
leads to the result that NMC  makes it harder to achieve
inflation with a given potential
that is known to be inflationary for 
$\xi=0$ \cite{Abbott81,FutamaseMaeda89,ALO90,Calgary}.
There are two main reasons for this difficulty:\\\\
1) The common attitude in the literature on nonminimally coupled
scalar
fields in inflation is that the coupling constant $\xi$ is a free
parameter to fine-tune at one's own will in order to solve problems of
the
inflationary scenario under consideration. The
fine-tuning of certain  parameters of inflation is reduced by
fine-tuning the extra parameter $\xi$ instead. For example, the
self-coupling
constant $ \lambda $ of the scalar field in the chaotic inflation 
potential $
V=\lambda \phi^4 $ is subject to the constraint $ \lambda < 10^{-12}
$ coming from the observational limits on the amplitude of
fluctuations in the  cosmic microwave background. This   
constraint makes the scenario uninteresting because 
the energy scale predicted by particle physics is much higher. The
constraint on $\lambda$ is reduced by
fine-tuning $\xi $ instead
\cite{SalopekBondBardeen89,FakirUnruh90a,FakirUnruh90b,
KolbSalopekTurner90,MakinoSasaki91}; while the
fine-tuning of $\xi $ is less drastic than that of the self-coupling
constant $\lambda$ by several orders of magnitude \cite{FakirUnruh90a}, 
one cannot be satisfied with the fact that NMC is
introduced {\em ad hoc} to improve the fine tuning problems (and still
does not completely cure them). A more rigorous approach consists in
studying the prescriptions for the value of $\xi $
given in the literature (which are summarized in Ref.~\cite{Faraoni96}) 
and the consequences of NMC for the known  inflationary
scenarios. The philosophy of this approach is that NMC is
often  unavoidable and the value of $\xi$ is not
arbitrary but is determined by the underlying physics. Once the value
of the coupling constant $\xi$ is predicted, one
does not have anymore the freedom to adjust its value and the
fine-tuning problems that may plague the inflationary
scenario reappear. Several inflationary scenarios
turn out to be theoretically inconsistent when one takes into account the
appropriate values of the coupling constant \cite{Faraoni96,Calgary}.
\\\\
2) Most of the
inflationary scenarios are built upon the
slow-roll approximation \cite{KolbTurner,Liddleetal94}, in which the
Einstein-Friedmann
dynamical equations are solved. It is more difficult to
achieve the slow
rolling of the scalar field when $\xi \neq 0$. In fact, an almost flat
section of the potential $V( \phi)$ gives slow rollover of $\phi$ when
$\xi =0$, but its shape is distorted by the
NMC term $\xi R \phi^2 /2$ in the Lagrangian density 
(\ref{Lagrangiandensity}). The extra term plays the role of an
effective mass term for the inflaton. The phenomenon was described
by Abbott \cite{Abbott81} in the new inflationary scenario with the
Ginzburg-Landau potential, by Futamase and Maeda \cite{FutamaseMaeda89}
in chaotic inflation, and by Fakir and Unruh
\cite{FakirUnruh92ApJ}; and the generalization 
to any slow roll inflationary potential is
straightforward \cite{Faraoni96}. This mechanism is quantitatively
discussed in Sec.~6.

How general are the previous conclusions~? They  hold for
particular inflationary scenarios, and the conclusion that it is {\em
always} more difficult to achieve a sufficient amount of inflation in the
presence of NMC is premature. In principle, it is possible that a
suitable scalar field
potential $V( \phi)$ be balanced by the NMC term $\xi
R \phi^2/2$ in the Lagrangian density 
(\ref{Lagrangiandensity}), thus producing an ``effective 
potential'' \cite{footnote4}  which
is inflationary and even gives a slow-roll regime. In this situation, 
NMC would make it easier to achieve inflation, thus opening
the possibility for a wider class of scalar field potentials to be
considered. This possibility is studied in this paper; the
discussion is kept as general as possible, without 
specifying a particular inflationary scenario until it is necessary.

In a previous paper \cite{Faraoni96}, the theoretical
consistency of the known inflationary scenarios was studied from the
point of view of
the theoretical prescriptions for the value of $ \xi$ and of the
fine-tuning of the parameters. 
Calculations of density perturbations with nonminimally coupled scalar
fields have been performed in Refs.
\cite{Hwang90,FakirHabib93,Kaiser95a,Kaiser95b,HwangNoh98,
KomatsuFutamase97,KomatsuFutamase99}, while observational constraints on
$\xi$ were derived in Refs.
\cite{Faraoni96,KomatsuFutamase97,HwangNoh98,
KomatsuFutamase99}.
Here instead we study the effect of    
NMC by analyzing the dynamical
equations for the scale factor of the
universe and the scalar field, without specifying the value of the
coupling constant $\xi$. Aspects of the physics of NMC which give 
rise to ambiguities in
the literature are also clarified.

Throughout this paper it is  assumed that gravity is described by
Einstein's theory with a scalar field as the only source of matter, as
described by the Lagrangian density (\ref{Lagrangiandensity}). Only in
Secs.~2 and 4.3 is the presence of a different kind of matter in addition
to
the scalar field allowed.

The plan of the paper is as follows: in Sec. 2 the possible ways of
writing the Einstein equations in the presence of NMC are discussed and
compared, together with the corresponding conservation laws and with
the issue of the effective gravitational constant. In Sec.~3 the
positivity
of the energy density of a nonminimally coupled scalar field is 
discussed in the context of cosmology. 
In Sec.~4 a necessary condition for the acceleration of a universe 
driven by a
nonminimally coupled scalar field is derived; this is relevant for both 
inflation and quintessence models based on scalar fields. The question of
whether
the acceleration can occur due to pure NMC without a
potential $V( \phi
)$ is answered.
In Sec.~5, scalar field potentials that are known to be inflationary for
$\xi=0$  are studied and it is shown that NMC spoils inflation rather
than helping it. In Sec.~6 the slow-roll approximation to inflation
with NMC and the attractor behavior of de Sitter solutions are  studied.
This is relevant for the calculation of density and gravitational wave
perturbations and, ultimately, for the comparison with observations of the
cosmic microwave background. Sec.~7 presents a discussion of conformal
transformation techniques
used in cosmology with NMC, while  Sec.~8 contains a discussion and the
conclusions.

\section{Field equations and conservation laws}

\setcounter{equation}{0}

When discussing nonminimally coupled scalar fields, many authors choose to
reason in terms
of an effective gravitational constant instead of keeping a
$\phi$-dependent term in the left hand side of the Einstein equations.
In this section this approach is discussed and compared with the
more conservative approaches using a $\phi$-independent
gravitational constant, and the corresponding conservation equations are
studied. The following discussion has early parallels, for
the special cases in Refs.~\cite{Deser70,Ellisalternatives} and a recent 
but incomplete one in Ref.~\cite{Madsen88}.

One begins from the action
\begin{eqnarray}  
S=S_g [ g_{cd} ] + S_{int}[ g_{cd} , \phi ] +S_{\phi}[ g_{cd},
\phi ] + S_{m} [ g_{cd},
\psi_m ] =& & \nonumber \\
= \int d^4 x \sqrt{-g} \left[ \left( \frac{1}{2\kappa}-\frac{\xi
\phi^2}{2} \right) R-\frac{1}{2} g^{ab} \nabla_a \phi
\nabla_b \phi  -V( \phi ) \right] + S_{m} [ g_{cd},
\psi_m ] \; ,
\label{S} & & 
\end{eqnarray}
where $\kappa \equiv 8\pi G $, $S_g=(2\kappa )^{-1} \int d^4x \sqrt{-g} R
$ is the purely
gravitational part of the action, $S_{int} =-\xi /2 \int d^4x \sqrt{-g}
\, R
\phi^2 $ is an explicit interaction term between the gravitational and
the $\phi$ fields, $S_{\phi}$ describes the  
purely material part of the action associated with the scalar field, and 
the remainder $S_m$ describes matter fields other than $\phi$,
collectively denoted by $\psi_m$.

The variation of the action (\ref{S}) with respect to $\phi$ leads to the
Klein-Gordon equation~(\ref{KG}). By varying Eq.~(\ref{S}) with respect to
$g_{ab}$ and using the well known formulas \cite{Wald}
\be
\delta \left( \sqrt{-g} \right) = -\frac{1}{2} \sqrt{-g} \, g_{ab} \,
\delta
g^{ab} \;
,
\ee
\be
\delta \left( \sqrt{-g} \, R \right) = \sqrt{-g} \left( R_{ab}-\frac{1}{2}
g_{ab} R \right) \delta g^{ab} \equiv \sqrt{-g} \, G_{ab} \, \delta g^{ab} 
\ee
(where $G_{ab}$ is the Einstein tensor), one obtains the
Einstein equations in the form
\be  \label{efe1}
\left( 1-\kappa \xi \phi^2 \right) G_{ab}=\kappa \left( \tilde{T}_{ab}[
\phi] + \tilde{T}_{ab}[ \psi_m ] \right) \equiv 
\kappa \tilde{T}^{(total)}_{ab} \; ,
\ee
where
\be    \label{Ttilde}
\tilde{T}_{ab}[ \phi ] =\nabla_a \phi \nabla_b \phi -\frac{1}{2} g_{ab}
\nabla^c \phi \nabla_c \phi -V g_{ab} 
+\xi \left[ g_{\mu\nu} \Box ( \phi^2) -\nabla_{\mu}\nabla_{\nu}
( \phi^2 ) \right] \; .
\ee
and
\be    
\tilde{T}_{ab}[ \psi_m ] =-\, \frac{2}{\sqrt{-g}} \frac{\delta S_m \left[
\psi_m , g_{cd} \right]}{\delta g^{ab}} 
\; .
\ee
One can also rewrite Eq.~(\ref{efe1}) by taking the factor $\kappa \xi
\phi^2 G_{ab}$ to the right hand side,
\be  \label{triangtriang}
G_{ab}=\kappa \tilde{\tilde{T}}_{ab} \; , 
\ee
where 
\be  \label{tdoubletilde}
\tilde{\tilde{T}}_{ab}=\tilde{T}_{ab}^{(total)}+\xi \phi^2 G_{ab} \; .
\ee
By taking a different approach, the coefficient
of the Ricci scalar in
the action (\ref{S}) can be written as $(16\pi G_{eff})^{-1}$, where 
\be   \label{Geff}
G_{eff} \equiv \frac{G}{1-8\pi G \xi \phi^2} 
\ee
is an effective, $\phi$-dependent, gravitational coupling. This way
of proceeding is analogous to the familiar identification  of the
Brans-Dicke scalar field $\phi_{BD}$ with the inverse of an effective
gravitational constant ($G_{\phi}=\phi_{BD}^{-1}$) in the gravitational
sector of the Brans-Dicke action
\be
S_{BD}=\int d^4 x \sqrt{-g} \left( \phi_{BD} R -\frac{\omega}{\phi_{BD}}
\nabla^a \phi_{BD} \nabla_a \phi_{BD} \right) \; .
\ee
By adopting this point of view in the case of a nonminimally coupled  
scalar field, one
divides Eq.~(\ref{efe1}) by the factor $1-\kappa \xi \phi^2 $
to obtain the Einstein equations in the form
\be  \label{efe2}
 G_{ab}=\kappa_{eff} \left( \tilde{T}_{ab} [ \phi ] + \tilde{T}_{ab}
[ \psi_m ] \right) =
\kappa_{eff} \, \tilde{T}^{(total)}_{ab} 
\ee
(where $\kappa_{eff} \equiv 8\pi G_{eff}$), which looks more familiar to
the relativist's eye.

The approach using the effective gravitational coupling (\ref{Geff}) has
been used to investigate the situation in which $G_{eff}=0$ and the
``antigravity'' regime corresponding to $G_{eff}<0$
\cite{antigravity}.

A third possibility is to use the form of the Einstein equations
\be  \label{efe3}
 G_{ab}=\kappa \left( T_{ab}[ \phi ] + T_{ab}[ \phi, \psi_m ] \right)
\equiv \kappa T_{ab}^{(total)} \; ,
\ee
where
\be    \label{T}
T_{ab} [ \phi ] \equiv \frac{1}{1-\kappa \xi \phi^2} \, \tilde{T}_{ab} [
\phi ]   \; ,
\ee
\be    
T_{ab} [ \phi ,\psi_m ] \equiv \frac{1}{1-\kappa \xi \phi^2} \,
\tilde{T}_{ab} [ \psi_m ]   \; ,
\ee
and the gravitational coupling is given by the true constant $G$.

If $\xi \leq 0$ the forms (\ref{efe1}), (\ref{triangtriang}), (\ref{efe2})
and
(\ref{efe3}) of the Einstein equations are all equivalent
(apart from the conservation of the corresponding stress-energy tensors,
which is discussed later). If instead $\xi >0$,
caution must be exercised to ensure that the factor $1-\kappa \xi \phi^2 $
by which Eq.~(\ref{efe1}) is divided does not vanish. The division by
$1-\kappa\xi \phi^2 $ used to write Eqs.~(\ref{efe2}) and (\ref{efe3})  
unavoidably introduces the two critical values of the scalar field
\be
\pm \phi_c=\pm \frac{m_{pl}}{\sqrt{8\pi \xi }}
\;\;\;\;\;\;\;\;\;\;\;\;\; (
\xi > 0) \; ,
\ee
which are barriers that the scalar field cannot cross. At $\phi=\pm \phi_c
$ the
effective gravitational coupling (\ref{Geff}), its gradient, and the
stress-energy tensor $T^{(total)}_{ab}$ in Eq.~(\ref{efe3}) diverge. 
Therefore, solutions of the field equations can only be obtained for which
$ \left| \phi \right| < \phi_c $ or $ \left| \phi \right| > \phi_c $ at
all times.  For $\xi>0$, one has obtained a {\em restricted}
form of the field equations and a restricted class of solutions; any
solution $\phi 
$ of the original theory described by Eq.~(\ref{efe1}) which crosses the
barriers $\pm \phi_c$ is lost in passing to the
picture of Eq.~(\ref{efe2}) or of Eq.~(\ref{efe3}).

Although the caveat on the division by the factor ($1-\kappa\xi\phi^2$)
looks
trivial, surprisingly it is missed in the literature on scalar
field cosmology with NMC, and the restricted range of validity of the
solutions goes unnoticed. In particular, investigations of the
coupled Einstein-Klein-Gordon equations  using dynamical
systems methods and aiming at determining generic solutions and
attractors, 
are put in jeopardy by the previous considerations if they employ the
form  
(\ref{efe2}) or (\ref{efe3}) of the Einstein equations.  For example, the
approach of Eq. (\ref{efe1}) is used in Ref.~\cite{ALO90}, which
makes correct statements on the general class of solutions of the field
equations with NMC, while the parallel treatment of Ref.
\cite{Barrosoetal92}) using Eq. (\ref{efe2}) cannot claim to study general
solutions. 

We proceed by discussing the conservation equations for $T_{ab} $ and
$\tilde{T}_{ab} $. The approach of Eq.~(\ref{efe1}) for $\xi >0 $ uses
a truly constant gravitational coupling $G$, but the field equations
(\ref{efe1}) do not guarantee covariant conservation of
$\tilde{T}_{ab}^{(total)
}$: in fact the contracted Bianchi identities $\nabla^b G_{ab}=0$ yield  
\be   \label{modifiedconservation}
\nabla^b \tilde{T}_{ab}^{(total)}=\frac{-2\kappa \xi \phi}{1-\kappa \xi
\phi^2} \, \tilde{T}_{ab}^{(total)} \nabla^b \phi 
\ee
when the denominator is nonvanishing.
The covariant divergence $\nabla^b \tilde{T}_{ab}^{(total)} $ vanishes
only for the trivial case $\phi =$const. and approximately vanishes in
regions of spacetime where $\phi$ is nearly constant. When the scalar
$\phi$ is the only source of gravity,
$\tilde{T}^{(total)}_{ab}=\tilde{T}_{ab}[ \phi ] $, the constancy of
$\phi$ corresponds to de Sitter solutions (if $\xi \leq 0$), with the
energy-momentum
tensor of quantum vacuum $\tilde{T}_{ab}=-V( \phi ) g_{ab} $ and equation
of state $P=-\rho$.

On the contrary,
in the approach based on Eq.~(\ref{efe3}) the relevant
stress-energy tensor $T_{ab}^{(total)} $ is covariantly conserved,
\be
\nabla^b T_{ab}^{(total)}=0 \; ,
\ee
as a consequence of the contracted Bianchi identities. This is probably
the reason why the approach based on
Eqs. (\ref{efe3}) has been preferred over alternative
formulations. However, the loss of generality in the solutions for
$\xi > 0 $ must be kept in mind.

To give an idea of how the conservation equation for ordinary
matter  
is modified by NMC, we consider the case of a dust fluid acting as the 
source 
of gravity together with the nonminimally coupled scalar field. When the
scalar identically vanishes, the equation $\nabla^b T_{ab}=0$ for the
stress-energy tensor $T_{ab} [ \psi_m ] =\rho u_a u_b $ (where $u^a$ is
the dust 
four-velocity) implies the geodesic equation for fluid particles
\be
u^b \nabla_b u^a =0 \; ,
\ee
and the conservation equation for the energy density $\rho$ 
\be
\frac{d\rho}{d\lambda}+\rho \nabla^b u_b = 0 \; ,
\ee
where $\lambda $ is an affine parameter along the geodesics. When the
nonminimally coupled scalar appears together with the dust,
Eq.~(\ref{modifiedconservation}) yields
\be
\left( \frac{d\rho}{d\lambda} +\rho \nabla^b u_b +\frac{2\kappa \xi \rho
\phi}{1-\kappa \xi \phi^2} \frac{d\phi}{d\lambda} \right) u_a + \rho
\frac{D u_a}{D\lambda}=0 \; ,
\ee
from which one derives again the geodesic equation $Du^a/ D\lambda \equiv
u^b \nabla_b u^a =0$ and the modified conservation equation
\be   \label{modified2}
\frac{d\rho}{d\lambda}+\rho \nabla^b u_b 
+\frac{2\kappa \xi  \phi\rho}{1-\kappa \xi \phi^2} \frac{d\phi}{d\lambda} 
= 0 \; .
\ee
The geodesic hypothesis \cite{Wald} is satisfied since test
particles move on
geodesics. In the weak field limit, the modified conservation equation
(\ref{modified2}) reduces to
\be  
\frac{\partial\rho}{\partial t}+ \vec{\nabla} \cdot\left( \rho \vec{v}
\right) 
+\frac{2\kappa \xi  \phi}{1-\kappa \xi \phi^2}
\left( \frac{\partial \phi}{\partial t} +\vec{\nabla}
\phi \cdot \vec{v} \right) \, \rho  = 0 \; .
\ee

Finally, we consider the approach using Eq.~(\ref{triangtriang}); it
employs the truly constant gravitational coupling $G$ and it 
guarantees that the
stress-energy tensor $\tilde{\tilde{T}}_{ab}$ is covariantly conserved,
\be
\nabla^b \tilde{\tilde{T}}_{ab} =0 \; ,
\ee
as can be deduced by using the contracted Bianchi identities and
Eq.~(\ref{triangtriang}).

\section{Positivity of the energy density}

\setcounter{equation}{0}

It is acknowledged \cite{Abreuetal94,Madsen88}  
that the energy density of a nonminimally coupled  scalar
field has a sign that depends on the particular solution $\phi $ and
on
the spacetime metric $g_{ab}$. This statement is easy to understand upon 
inspection of the rather complicated expression (\ref{density}) for the
energy
density of a nonminimally coupled  scalar in a
Friedmann-Lemaitre-Robertson-Walker (FLRW) 
universe. Since it is difficult or impossible to establish {\em a priori}
the sign of $ \rho$, the minimal physical requirement $\rho \geq 0 $ has
to be checked {\em a posteriori} for the known solutions of the field
equations. 

In this section we limit the discussion to homogeneous and isotropic
cosmologies; scalar fields are extremely important in this context, due to
their role as inflaton, dark matter, and quintessence. In this context,
it is possible to improve substantially on the subject of the positivity
of the energy density.

In a spatially flat or closed (curvature index $K \geq 0$) FLRW
universe dominated by a scalar field, the solution $\left( a(t) , \phi (t)
\right) $ of the field equations satisfies the Hamiltonian constraint
\be
\left( \frac{\dot{a}}{a} \right)^2=\frac{\kappa }{3} \rho -\frac{K}{a^2} 
\; ,
\ee
which follows from the field equations in the form (\ref{triangtriang})
and
(\ref{tdoubletilde}),
and  from which it is immediate to deduce that the energy density
$\rho$ is {\em always} non-negative {\em for a solution of the Einstein
equations}, in spite of the complication of the expression
of $\rho$ in terms of $\phi$,
$\dot{\phi}$, $a$ and $H=\dot{a}/a$ (see for example
Eq.~(\ref{densityy})).

Note that the different forms of the field equations considered in the
previous section lead to different stress-energy tensors, and therefore to
different definitions of energy density of the scalar field. Hence,
by using field equations different from
(\ref{triangtriang}) and
(\ref{tdoubletilde}) one is not able to draw conclusions on the sign of
$\rho$.

\section{Necessary conditions for the acceleration of the universe}

\setcounter{equation}{0}

In the rest of this paper we specialize our considerations  to cosmology.
In this section, we study a necessary condition for the universe to
accelerate when a
nonminimally coupled scalar field is the dominant source of gravity. This
is relevant for both inflation and quintessence models based on scalar
fields with NMC. While such inflationary models are well known
\cite{FakirUnruh90a,Barrosoetal92,KomatsuFutamase97,KomatsuFutamase99,
HwangNoh98,FutamaseTanaka,BassettLiberati,Amendola99},
the use of scalar fields with NMC as dark matter
\cite{LTMI,Morikawa} and, more recently, as quintessence models
\cite{Chiba,Uzan,PBM} is less known. The necessary condition for cosmic
acceleration derived here is 
applied in the following sections.

The study of necessary conditions for the acceleration of the universe  
enables one to determine
whether NMC helps, or
makes it difficult to achieve inflation for a given scalar field
potential, in comparison with the corresponding situation for minimal
coupling.  
  
We begin by considering the spatially flat Einstein-de Sitter
universe with line element  
\be  \label{F1}
ds^2=-dt^2+a^2(t) \left( dx^2+dy^2+dz^2 \right) 
\ee
in comoving coordinates $\left( t,x,y,z \right)$.
In this section we adopt the form (\ref{efe3}) of the Einstein field
equations 
keeping in mind the caveat of Sec.~2; the Einstein-Friedmann equations are
\begin{equation}    \label{F2}
\frac{\ddot{a}}{a}= - \, \frac{\kappa}{6} ( \rho +3P ) \; ,
\end{equation}
\begin{equation}
H^2 \equiv \left( \frac{\dot{a}}{a} \right)^2=  \frac{\kappa}{3} \rho  \;
,
\end{equation}
where an overdot denotes differentiation with respect to the comoving time
$t$. 
The energy density $\rho$ and
pressure $P$ of the scalar field are given by the diagonal components of
the stress-energy tensor $T_{ab}[ \phi ]  $ of Eq. (\ref{T}):
\begin{equation} \label{density}
\rho=\left( 1-\kappa \xi \phi^2 \right)^{-1} \left[ \frac{(
\dot{\phi})^2}{2}+V( \phi)+6\xi H \phi \dot{\phi}\right] \; ,
\end{equation}
\begin{equation}                  \label{pressure}
P=\left( 1-\kappa \xi \phi^2 \right)^{-1} \left[ \left( \frac{1}{2}-2\xi 
\right) 
\dot{\phi}^2-V( \phi)-2\xi \phi \ddot{\phi}-4\xi H \phi \dot{\phi} \right]
\; .
\end{equation}             
Equations (\ref{F1}), (\ref{density}) and (\ref{pressure}) yield, in the
case of minimal coupling ($\xi=0$),
\begin{equation}  \label{ddota}
\frac{\ddot{a}}{a}= - \, \frac{\kappa}{3} \left( {\dot{\phi}}^2 -V
\right)
\;  .
\end{equation}

An inflationary era in the evolution of the universe includes as an
essential feature an accelerated expansion, $\ddot{a}>0$ (of course, other
ingredients are required for a successful inflationary scenario: a
natural mechanism of entry into inflation, a sufficient amount of
expansion, a graceful exit mechanism, acceptable scalar and tensor
perturbations, etc). It is clear
from Eq.~(\ref{ddota}) that when $\xi=0$ a necessary (but
not sufficient) condition for acceleration is given by $V>0$.
It is useful to keep in mind that the slow-roll approximation used to
solve the
equations of inflation, corresponds to the dominance of
the scalar field potential energy density $V( \phi )$ over its kinetic
energy density, $ V( \phi ) >> \dot{\phi}^2 /2$. In the slow-roll
approximation for $\xi=0$, $\rho \approx V$ and hence the necessary
condition for acceleration $V>0$ reduces to the minimal requirement of the
positivity of the scalar field energy density, and of the existence of
real solutions of Eq.~(\ref{F2}).

What is the necessary condition for acceleration analog to $V>0$ when
$\xi \neq 0$ ? The first
part of this  section is devoted to answering this question.

By definition, acceleration corresponds to the condition $\rho +3P<0$;
upon
use of Eqs. (\ref{density}) and (\ref{pressure}), this inequality is
equivalent to
\begin{equation}
(1-3\xi ) {\dot{\phi}}^2 -V-3\xi \phi \left( \ddot{\phi}+H \dot{\phi}
\right)<0
\; .
\end{equation}
The Klein-Gordon equation (\ref{KG}), which takes the form
\begin{equation}  \label{KG2}
\ddot{\phi}+3H\dot{\phi} +\xi R \phi +\frac{dV}{d\phi}=0 \; ,
\end{equation}
is then used to substitute for $\ddot{\phi}$, obtaining
\begin{equation}       \label{inequality}
(1-3\xi ) {\dot{\phi}}^2 -V+3\xi^2 R\phi^2 + 6\xi H \phi \dot{\phi} +3\xi
\phi 
\frac{dV}{d\phi} <0   \; ,
\end{equation}
and Eq.~(\ref{density}) can be used to rewrite the definition of
cosmic acceleration (\ref{inequality}) as 
\begin{equation}
x\equiv \left( 1-\kappa \xi \phi^2 \right) \rho -2V+\dot{\phi}^2
\left( \frac{1}{2}-3\xi \right)+3\xi^2 R\phi^2+3\xi \phi \,
\frac{dV}{d\phi} <0 \;.
\end{equation}
To proceed, one assumes the weak energy condition $\rho \geq 0$. As a
result of 
the difficulty of handling the dynamical equations analytically 
when $ \xi \neq 0$, in the rest of this section we restrict
ourselves to values of the coupling constant $\xi \leq 1/6$. Albeit
limited, this semi-infinite range covers many of the prescriptions for
the value of $\xi$ given in the literature \cite{Faraoni96}. One then has 
$ -2V+3\xi \phi dV/d\phi \leq x <0 $ and a necessary condition 
for cosmic acceleration to occur when $\xi \leq 1/6$ is 
\begin{equation}    \label{abc}
V-\frac{3\xi}{2}\phi \, \frac{dV}{d\phi} >0 \; .
\end{equation}
Eq.~(\ref{abc}) reduces to the well known necessary condition
for acceleration $V>0$ for minimal coupling.

Unfortunately, the necessary and sufficient condition for acceleration 
(\ref{inequality}) is not very useful in general
because different terms, which depend on the solution $\left( a(t) , \phi
(t) \right) $ and have opposite signs can balance one another and
hamper a general analysis of the problem. In practice, one is compelled to
adopt one of the specific forms of $V( \phi)$ considered in the
literature and solve the equations for $a(t)$ and $\phi (t)$ for 
specific examples. However, a few considerations of general character can
still be given.

First, we take the point of view that a potential $V( \phi )$ is
given, for example by a particle physics theory, and we study the effect
of introducing NMC in the field equations. The
discussion is kept as general as possible, without specifying the value
of $ \xi$.

Keeping in mind the necessary condition (\ref{abc}) for acceleration of
the universe in the
$\xi =0$ case, consider an even potential $V( \phi )=V( -\phi )$ which
is increasing for $ \phi > 0$. This is the case, e.g., of a pure mass term
$ m^2 \phi^2 /2$, or of the quartic potential $ V=\lambda \phi^4$, or of
their combination $V( \phi ) = m^2 \phi^2 /2 + \lambda \phi^4 + V_0$,
where $ V_0 $ is
constant. For
$ 0 < \xi < 1/6$, one has $\xi \phi dV/d\phi >0 $ and it is harder to
satisfy the necessary
condition (\ref{abc}) for acceleration than in the minimal coupling case.
Hence one can
say that, for this class of potentials, it is harder to achieve
acceleration of the universe, and hence inflation.
If instead $ \xi <0 $, the necessary condition for cosmic acceleration is
more
easily satisfied than in the $\xi=0 $ case,
but one is not entitled to say that with NMC it is easier to achieve
inflation (because a necessary, and not a sufficient
condition for acceleration is considered).

Let us consider now an even potential $V( \phi )=V( -\phi ) $ such that 
$ dV/d\phi < 0 $ for $ \phi >0$. This is the case, e.g., of
the Ginzburg-Landau potential $V( \phi) =\lambda \left( \phi^2 -v^2
\right)^2 $ for $0 < \phi < v$, or of an inverted harmonic oscillator
potential \cite{LiddleLyth}, which approximates the potential for natural
inflation 
\be
V_{ni}( \phi )= \lambda^4 \left[ 1+ \cos \left( \frac{ \phi}{f} \right) 
\right]
\ee
around its maximum at $ \phi =0 $. For $ 0< \xi \leq 1/6 $, it is easier
to satisfy the necessary condition (\ref{abc}) for acceleration 
when $\xi \neq 0$ than when $\xi =0$ but, again,
this does not allow one to conclude that the universe actually
accelerates its expansion. If
$ \xi <0 $ instead, it is harder to achieve acceleration
than in the $\xi = 0 $ case.

The inequality (\ref{abc}) can be read in a different way: assume, for
simplicity, that $V>0$ and $\phi >0$ (it is straightforward to generalize
to the case in which $V$ or $\phi$, or both, are negative). Then, if $ 0
\leq \xi \leq 1/6$, (\ref{abc}) is equivalent to 
\be
\frac{d}{d\phi} \, \left\{ \ln \left[ \frac{V}{V_0} \, \left(
\frac{\phi_0}{\phi} \right)^{\frac{2}{3\xi}}\right] \right\} <0 \; ,
\ee
where $V_0$ and $\phi_0$ are arbitrary (positive) constants. As a result
of  the
fact that the logarithm is a monotonically increasing function of its
argument, the necessary condition for cosmic acceleration (\ref{abc})
amounts to require that the potential $V( \phi )$ grows with $\phi$ slower
than the power-law potential $V_{crit}( \phi ) \equiv V_0 \left(
\phi/\phi_0 \right)^{\frac{2}{3\xi}}$. If instead $\xi <0$, the necessary
condition for cosmic acceleration amounts to require that $V$ grows {\em
faster} than $V_{crit} ( \phi ) $ as $\phi$ increases.  This criterion
is further developed in Sec.~4.3.

\subsection{No acceleration without a scalar field potential}

Taking to the extreme the possibility of a balance between the potential
$V( \phi)$ and the term $\xi R \phi^2/2$ in (\ref{Lagrangiandensity}), the
question arises of whether it is possible to obtain acceleration of the
universe 
with a free, massless scalar field with no cosmological constant 
(i.e. $V=0$) for suitable values of the
coupling constant $\xi$. In particular, we are interested to the case of 
strong coupling $| \xi |
>> 1 $, which has been considered many times in the literature
\cite{SalopekBondBardeen89,FakirUnruh90a,Morikawa,LTMI,BassettLiberati,
HwangNoh98,Chiba}.

Inflation driven by a pure NMC term  turns out to be impossible for
negative values of $\xi$. In fact, by assuming that the expansion of the 
universe is accelerated and
that $V=0$,
 Eq.~(\ref{inequality}) and the expression of the Ricci curvature in an
Einstein-de Sitter space
\begin{equation}         \label{Ricci}
R=6 \left( \frac{\ddot{a}}{a}+\frac{{\dot{a}}^2}{a^2} \right) >6H^2 
\end{equation}
yield, for $\xi<0$,
\begin{equation}       
(1-3\xi ) {\dot{\phi}}^2 +3\xi^2 R\phi^2 + 6\xi H \phi \dot{\phi} \geq
\left(
\dot{\phi}+3\xi H \phi \right)^2 \geq 0  \; ,
\end{equation}
thus contraddicting (\ref{inequality}). Therefore, the combined assumptions
$\ddot{a}>0$ and $V=0$ lead to an absurdity. The previous analysis 
fails to yield conclusions when $\xi>0$ because terms of different signs
can balance in the left hand side of (\ref{inequality}). The previous
reasoning does not make use of the weak energy condition.

The discussion can easily be extended to values
of $\xi$ in the range $0<\xi \leq 1/6$ by using an independent argument:
by rewriting (\ref{inequality}) as
\begin{equation}        \label{100}
\dot{\phi}^2 \left( \frac{1}{2}-3\xi  \right)< - \left[ 3\xi^2 R \phi^2
+\left( 1-\kappa \xi \phi^2 \right) \rho \right]   \; ,
\end{equation}
and using $\rho \geq 0$, one concludes that the right hand side is
negative when the cosmic expansion accelerates and that (\ref{100}) can be
satisfied only if the
term on the left hand side is negative, i.e. if $\xi >1/6$, which
contraddicts the assumptions. Therefore,\\\\ 
{\em  for $ \xi \leq 1/6$, the NMC term alone cannot
act as an effective potential to provide acceleration of the
universe.}\\\\ 
A further argument (again for $\xi \leq 1/6$) 
consists in noting that the necessary condition for
acceleration (\ref{abc}) is not satisfied if $V( \phi)$ vanishes
identically. 
Unfortunately no conclusion can be obtained analytically when $\xi > 1/6$.

\subsection{Negative potentials}

In the usual studies of inflation and quintessence with $\xi =0$, only
positive scalar
field potentials are considered. The reason is easy to
understand: in the slow-rollover approximation to
inflation $V>>\dot{\phi}^2/2$, $
\rho \approx V $, and $V>0$ corresponds to $ \rho >0$, a minimal
requirement. However, this is no longer true when
$\xi \neq 0 $ and $\rho$ is given by the more complicated expression
(\ref{density}).
Indeed, negative scalar field potentials have been considered in the
literature on NMC, in inflation \cite{Abreuetal94} or in 
other contexts \cite{BechmannLechtenfeld95,RatraPeebles88,Hosotani85},

The question of whether the positive
term $ \xi R \phi^2 /2$ can balance a negative $V( \phi ) $ arises. 
In the toy model of Ref.~\cite{Bayinetal94} a negative potential $V$ is 
balanced by the coupling term $\xi R \phi^2 /2 $ in such a way that
inflation is achieved: there, a closed
FLRW universe dominated by a conformally coupled scalar
field is investigated in Einstein's gravity. By assuming
the 
equation of state $P=( \gamma-1) \rho$, the 
potential deemed necessary for inflation is derived numerically for small 
values of the constant $\gamma$; the resulting $V( \phi )$  is
significantly 
different from the corresponding potential derived analytically in 
Ref.~\cite{StarkovichCooperstock92} for the case $\xi=0$ and for the same
values of $\gamma$.

Mathematically, the possibility of a negative $ V( \phi )$ which is
inflationary in the presence of NMC extends the range of potentials 
explored so far, but the meaning of a negative scalar field potential $
V( \phi ) $ remains unclear and the latter is probably unpalatable to most
particle physicists.

\subsection{Quintessence models}

In quintessence models based on a nonminimally coupled scalar field, the
energy density of the latter is beginning to dominate the dynamics 
of the universe, but there is also ordinary matter with energy density
$\rho_m \propto a^{-3} $ and vanishing pressure $P_m=0$.
Eq.~(\ref{density}) is modified according to 
\begin{equation} \label{totaldensity}
\rho=\left( 1-\kappa \xi \phi^2 \right)^{-1} \left[ \rho_m + \frac{(
\dot{\phi})^2}{2}+V( \phi)+6\xi H \phi \dot{\phi}\right]
=\frac{\rho_m}{1-\kappa \xi \phi^2} + \rho_{\phi} \; ,
\end{equation}
where $V( \phi ) $ is an appropriate quintessential potential. 
The necessary and sufficient condition for the acceleration of the
universe $\rho + 3P <0 $ is written as 
\begin{equation}
y \equiv \frac{\rho_m}{2}+  
\left( 1-3\xi \right) \dot{\phi}^2 -V +6\xi H \phi \dot{\phi} +3\xi^2
R\phi^2+3\xi \phi \, \frac{dV}{d\phi} <0 \; ,
\end{equation}
where the Klein-Gordon equation (\ref{KG2}) has been used to substitute
for $\ddot{\phi}$. As before, one obtains a necessary condition for the
accelerated expansion of the universe by rewriting the quantity $y$ as
\begin{equation}
y = \frac{\rho_m}{2} + \left( 1-\kappa \xi \phi^2 \right) \rho_{\phi} 
-2V+\dot{\phi}^2
\left( \frac{1}{2}-3\xi \right)+3\xi^2 R\phi^2+3 \xi \phi \,
\frac{dV}{d\phi} <0 \; ,
\end{equation}
and by assuming that $\rho_m$ and $\rho_{\phi}$ are  non-negative; one
obtains again Eq.~(\ref{abc}) as a necessary condition for the
acceleration of a 
universe in which quintessence is modelled by a nonminimally
coupled scalar field, in the additional presence of ordinary matter.

The analysis can be refined by noting that, when quintessence dominates,
$\rho_{\phi} \approx V$. By introducing the matter and scalar field energy
densities measured in units of the critical density $\rho_c$
(respectively, $\Omega_m=\rho_m/\rho_c $ and
$\Omega_{\phi}=\rho_{\phi}/\rho_c $), one has $\rho_m \simeq V
\, \Omega_m/\Omega_{\phi}$ and
\be
y=\left( -1 +\frac{\Omega_m}{2\Omega_{\phi}}-\kappa\xi \phi^2 \right) V+
\left( \frac{1}{2} -3\xi \right) \dot{\phi}^2 +3\xi^2 R \phi^2 +3\xi \phi
\frac{dV}{d\phi} <0 \; .
\ee
For $\xi \leq 1/6$ one has
\be
\left( -1 +\left. \frac{\Omega_m}{2\Omega_{\phi}}\right|_0  -\kappa\xi
\phi^2 \right) V  +3\xi \phi \frac{dV}{d\phi} \leq y <0 \; ,
\ee
where the ratio $\Omega_m/\Omega_{\phi}$ has been approximated by its
present value (which is correct at least around the present epoch). By
assuming again, for simplicity, that $V$ and $\phi$ are positive, the
necessary condition for quintessential inflation for $\xi \leq 1/6$ is
\be
\frac{d}{d\phi} \left\{ \ln \left[ 
\frac{V}{V_0} \left( \frac{\phi_0}{\phi} \right)^{\alpha} \mbox{exp}\left(
-\frac{\kappa}{6} \phi^2 \right) \right] \right\} <0 \; ,
\ee
where $V_0$ and $\phi_0$ are constants and 
\be
\alpha = \left( 1- \left. \frac{\Omega_m}{2\Omega_{\phi}} \right|_0 
\right) \, \frac{1}{3\xi} \; .
\ee
Then, to have quintessential expansion with nonminimal coupling and $0 <
\xi \leq 1/6 $, one needs a potential $V( \phi )$ that does not grow with
$\phi$ faster than the function
\be  \label{Cdiphi}
C( \phi ) = V_0 \left( \frac{\phi}{\phi_0} \right)^{\alpha} \exp 
\left( \frac{\kappa}{6} \phi^2 \right) \; .
\ee
If instead $\xi <0$, $V( \phi )$ must grow faster than $C( \phi )$. 
The necessary conditions for cosmic acceleration in a
quintessence-dominated universe are useful for future reference in studies
of quintessence models with NMC.

\section{Fixing the scalar field potential}

\setcounter{equation}{0}

As a result of the complication of the coupled
Einstein-Klein-Gordon
equations when $ \xi \neq 0$, general analytical considerations on the
occurrence of inflation
with nonminimally coupled scalar fields are necessarily quite limited, as
seen in Sec.
2. However, one can (at least partially)  answer the following meaningful
question:\\
{\em is it
harder or easier to achieve acceleration of the universe with NMC for
the potentials that are known to be inflationary in the minimal coupling
case~?} 
\\ Since in many situations these potentials are motivated by a
high energy physics theory, they are of special interest.
In order to appreciate the effect of the inclusion of a NMC term in a
given inflationary scenario, we study some exact
solutions for popular inflationary potentials, and the necessary condition
(\ref{abc}) for the occurrence of inflation.

\subsection{$ V=0$}

In order to illustrate the qualitative difference between minimal and
nonminimal coupling, it is sufficient to compare the solution for
$V=0$, $ \xi=0$ with the corresponding solution for the special value
of the NMC coupling constant $\xi =1/6$. For minimal coupling, one
has    
the stiff equation of state $ P=\rho$, and the scale factor $ a(t)= a_0 \, 
t^{1/3}$, as can be deduced by the inspection of Eqs.~(\ref{density}) and
(\ref{pressure}) (we exclude the trivial case of Minkowski space). In the
$ V=0$, $\xi =1/6$ case, the Klein-Gordon
equation is
conformally invariant, corresponding to the vanishing of the trace of $ 
T_{ab} [ \phi ]$, to the radiation equation of
state $P=\rho/3$, and to the scale factor  $a(t)=a_0 \, t^{1/2} $. This is
in agreement with the fact that there are no accelerated universe  
solutions
for $V=0$ and any value of $\xi$, because the necessary condition 
(\ref{abc}) cannot be satisfied in this case.

\subsection{$ V=V_0 =$constant}

For $ \xi =0$ a constant potential can be regarded as a
cosmological constant in the Einstein equations. Viceversa, a $ \Lambda
$-term in the Einstein equations,
\be
G_{ab}=\Lambda g_{ab} + \kappa T_{ab} \; ,
\ee
can be incorporated into the scalar field potential by means of the
substitution 
\be
V( \phi ) \rightarrow V( \phi ) +\frac{\Lambda}{\kappa}  \; ;
\ee
this is true subject to the condition that the scalar field be constant.

The equivalence between cosmological constant and constant scalar field
potential no longer holds when $ \xi \neq 0$ and the form 
 (\ref{efe3}) of the Einstein equations is used. In fact,
in this case the addition of a cosmological constant term to the left
hand side of the  Einstein equations (\ref{efe3}),
\be
\kappa T_{ab} \rightarrow \kappa T_{ab} + \Lambda g_{ab} 
\ee
is equivalent to the substitution 
\be
V( \phi ) \rightarrow V( \phi ) +\frac{\Lambda}{\kappa} \left( 1-\kappa 
\xi \phi^2 \right) =V ( \phi ) + V_1 ( \phi ) \; .
\ee
The extra piece $ V_1 ( \phi )=\frac{\Lambda }{\kappa} \left(
1-\kappa\xi \phi^2 \right) $ in the
potential does not correspond to a mere shift in the
potential energy density (usually identified with the vacuum energy), but
it also adds a self-interaction with the shape of an inverted harmonic
oscillator. This is an example of how different things are when $ \xi 
$ is allowed to be different from zero, and testifies to the difference
between  the 
physical interpretations associated with the different ways of writing the
field equations discussed in Sec.~2. When $ \xi \neq 0 $, a constant
potential cannot be interpreted as the vacuum energy density coming from
the left hand side of the Einstein equations. 

In the case $ V= V_0$, the necessary condition (\ref{abc}) for
cosmic acceleration when $ \xi \leq 1/6$
coincides with the corresponding condition for minimal coupling, $
V=\Lambda/\kappa >0$. A negative $ V_0 $ does not give rise to
acceleration of the universe and hence to inflation. While for $ \xi
=0$ a negative $ \Lambda $ violates the
weak energy condition, this may no longer be true for $ \xi \neq 0$. The
necessary and sufficient condition for acceleration when $ \xi = 0$
is $ \Lambda > \kappa \dot{\phi}^2 /2 $, which recalls the
slow-roll condition.

The $ \xi =0$, $V=$const. case itself deserves comment. In 
this case, the Einstein-Friedmann equations with $ V= \Lambda/\kappa $
have the
familiar de Sitter solution (historically, the prototype of
inflation)
\be   \label{ES1}
a(t)=a_0 \exp( Ht ) \; , \;\;\;\;\;\, \;\;\;\;\;\;\; \dot{H}=0 \; , 
 \;\;\;\;\;\, \dot{\phi}=0 \; ,
\ee
corresponding to the vacuum equation of state $ P=-\rho $. In addition one
has, for $ \Lambda \neq 0$, the exact solution
\be   \label{ES2}
a(t)= a_1 \left[ \sinh \left(
\sqrt{3 \Lambda} \,\, t \right) \right]^{1/3} \; ,
\ee
corresponding to the non-constant scalar field
\be   \label{ES3}
\phi (t)= \pm \sqrt{ \frac{2}{3\kappa}} \, \ln \left[ \tanh \left(
\frac{\sqrt{3\Lambda}}{2} \, \, t \right) \right] + \phi_0 \; ,
\ee
where $a_1$ and $\phi_0 $ are integration constants. The latter solution
is asymptotic to (\ref{ES1}) at late times $ t \rightarrow + \infty $, in
agreement with the cosmic no-hair theorems \cite{KolbTurner}, 
but it exhibits a big-bang singularity at $t=0$ (where $a(t) \simeq
t^{1/3}$), while the FLRW universe
described by the de Sitter solution (\ref{ES1}) has
been expanding forever. In addition, the solution (\ref{ES2}) and 
(\ref{ES3})
corresponds to an effective equation of state that changes with time, and
interpolates between the two extremes $ P=\rho $ (``stiff'' equation of
state) at early times $t\rightarrow 0 $ and the vacuum equation of state $
P=-\rho $ at late times.
The exact solution 
(\ref{ES2}) and (\ref{ES3}) tells us two things (note that we are not even 
talking about the more complicated NMC  in this
example):\\\\
1) Contrarily to naive statements found in the literature, fixing the
scalar
field potential $ V( \phi ) $ does not fix the equation of state,
and, therefore, the scale factor $a(t)$.\\\\
2) A given potential $ V( \phi ) $ may correspond to very different
equations of state, depending on the solution   $   \left( g_{ab}, \phi
\right) $ of the field equations.\\

To conclude this subsection, we note that one can {\em impose} that the
solution (\ref{ES1}) hold (in the spirit, e.g., of Ref.~\cite{LTMI}); in
this case, if $ \xi =0$, the Klein-Gordon \cite{footnote5} equation
implies that $ \left. 
dV/d\phi \right|_{\phi_0} =0$. If instead $
\xi \neq 0$, by imposing that the solution (\ref{ES1}) hold, the
Klein-Gordon equation implies that $ \left. dV/d\phi 
\right|_{\phi_0}=-12
\xi H^2 \phi_0 $. A positive linear potential $V=\lambda \phi $ achieves
de Sitter expansion
with constant $\phi$ if $\xi <0$ and $H^2 =\lambda / ( 12 | \xi | ) $.

\subsection{$ V=m^2 \phi^2 /2 $}

A pure mass term is perhaps the most natural ``potential'' for a scalar
field, and an example of the class of even potentials for which
$ \phi dV/ d\phi > 0 $ considered in Sec.~4. For $ \xi =0$, it
corresponds to chaotic inflation \cite{Linde}, while for $ \xi <0 $, it
can still generate inflation. For example, the exponentially expanding
solution
\be
H=H_*=\frac{m}{(12 | \xi |)^{1/2}} \; , \;\;\;\; 
\phi = \phi_*=\frac{1}{(\kappa | \xi |)^{1/2}} 
\ee
has been studied, not in relation to the early 
universe, but as the description of short periods of unstable
exponential expansion of the universe which occur after the star
formation epoch, well into the matter dominated era \cite{LTMI}. The fact 
that the Ricci curvature $R$ is constant for this particular solution 
makes this case particularly suitable for the interpretation of the $ \xi
R
\phi^2 /2 $ term in
the Lagrangian density as a negative mass term, which balances the
intrinsic mass term $ m^2 \phi^2 /2 $ in the potential, thus conspiring
to give
a vanishing effective mass $ m_{eff}=\left( m^2 -|\xi | R \right)^{1/2} $.
However, the
so called late time mild inflationary scenario \cite{LTMI} based on the
relation $m_{eff}=0$ is
unphysical, as is best seen by studying the scalar wave tails in the
corresponding spacetime \cite{FaraoniGunzig98}. In fact, the scenario 
corresponds to a spacetime in which a {\em massive} scalar field
propagates sharply on the light cone at every point; this occurs because
the usual tail due to the intrinsic mass $m$ is cancelled by 
a second tail term describing the backscattering of the
$\phi$ waves off the background curvature of
spacetime \cite{FaraoniGunzig98}.

For $\xi >0 $, the nonminimally coupled scalar field has been studied
by Morikawa \cite{Morikawa}, who found no inflation. The $\xi >0 $ case
was studied 
in order to explain the reported periodicity in the redshift of galaxies
\cite{BEKS}. If the model
was correct, the parameter $\xi $ could be determined directly from
astronomical observations, and it would provide information on whether
general relativity is the correct theory of gravity \cite{FaraoniGRG}.
However, the prevailing opinion among astronomers is that the reported
periodicity in galactic redshifts is not genuine, but is an artifact of
the statistics used to analyze the astronomical data. The nonminimally
coupled, massive, scalar field model may however be resurrected in the
future in
conjunction with the more recent claims of redshift periodicities for
large scale structures \cite{Einastoetal}.

From the point of view of this paper, the
introduction of NMC  destroys the acceleration of the cosmic
expansion for large
positive values of $ \xi $ when $V( \phi) =m^2 \phi^2 /2$. This is
relevant
since we were not able to draw
conclusions for $ \xi > 1/6 $ in Sec.~4.  

\subsection{Quartic potential}

The potential $V= \lambda \phi^4 $ corresponds to chaotic inflation for $
\xi=0 $.   When $\xi \neq 0$ we limit ourselves to
consider the case of conformal coupling. For $\xi=1/6$, the Klein-Gordon
equation (\ref{KG2}) is conformally invariant, corresponding to the
vanishing of the trace $ T=\rho -3P $ of the
scalar field stress-energy tensor, to the radiation equation of state $
P=\rho /3 $ and to the non-inflationary expansion law $ a(t)
\propto t^{1/2} $. The introduction of conformal coupling destroys the
acceleration occurring in the minimal coupling case for the same
potential; however, accelerated solutions can be recovered by breaking the
conformal symmetry with the introduction of a mass for the scalar or of a 
cosmological constant (which, in this respect, behaves in the same manner 
\cite{MadsenGRG}). Exact
accelerating and non-accelerating solutions corresponding to integrability
of the dynamical system for the 
potential $V=\Lambda+m^2\phi^2/2 +\lambda \phi^4$ are presented in
Ref.~\cite{TALEV} for special sets of the parameters $\left(
\Lambda, m, \lambda \right)$.

\subsection{$V= \lambda \phi^n $}

In general, the necessary condition for cosmic acceleration (\ref{abc})
depends on 
the particular solution of the Klein-Gordon equation, which is not 
known {\em a priori}. However, this dependence disappears for power-law
potentials.  This case contains those of the previous subsections and also
the potential $ V \propto \phi^{-|\beta |}$, which approximates the
potential for intermediate inflation \cite{BarrowSaich90} and has been
used in quintessence models
\cite{Steinhardtetal,Zlatevetal,RatraPeebles88,Chiba,Uzan}.
The previous examples can be extended to the case of a potential
proportional to an even power of the scalar field, which is associated to
chaotic inflation for $ \xi =0$. The
necessary condition (\ref{abc}) for the occurrence of accelerated cosmic
expansion then
becomes
\be 
\lambda \left( 1-\frac{3n\xi}{2} \right) >0
\ee
when $ \xi \leq 1/6$. Under the assumption $ \lambda >0 $ corresponding
to a positive scalar field potential, the necessary
condition
(\ref{abc}) for acceleration of the cosmic expansion fails to be
satisfied when $ \xi \geq 2/3n$, independently of the solution $ \phi $
and of the initial conditions ($ \phi_0, \dot{\phi}_0 $). This is
interesting for $ n \geq 4$. Hence, also in this case, NMC destroys
acceleration in the range of values of $\xi$ $\left( 2/3n \right. , \left. 
1/6 \right]$.

The potential $ V=\lambda \phi^n $ with $n >6$ gives rise to
power-law inflation $a=a_0 t^p $, where
\be
p= \frac{ 1+ (n-10) \xi }{(n-4)(n-6) | \xi |}
\ee
(\cite{FutamaseMaeda89}; see also Ref. \cite{Abreuetal94} for the case $
\xi =1/6$). The universe is accelerating if $p>1$. The range
of
values
$ 6< n  \leq 10 $ is interesting for superstring theories
\cite{Shafi93,LiddleLyth}; however, the scenario is fine-tuned for $
\xi
>0$ \cite{FutamaseMaeda89}. For $ \xi < 0 $ the
solution is accelerating only if $ 6< n < 4+2 \sqrt{3} \simeq 7.464 $. 

\subsection{Exponential potential}

The potential $ V=V_0 \exp \left( -\sqrt{2\kappa /p} \, \phi \right)
$
is  associated to power-law inflation $a=a_0 \, t^p $ when $ \xi =0$ and
$\phi > 0$. An exponential potential is the
fingerprint of theories which are reformulated in the Einstein conformal
frame by means of a suitable conformal transformation of a theory
previously set in the Jordan conformal frame (see Ref.
\cite{FaraoniGunzigNardone99} for an explanation of this terminology and
for the relevant formulas). This class of theories include Kaluza-Klein
and higher-dimensional theories with compactified extra dimensions,
scalar-tensor theories of gravity, higher derivative and string
theories \cite{FaraoniGunzigNardone99}. In this case, the low energy
prediction for the coupling constant yields the value $\xi =0$
\cite{Faraoni96}.
Nevertheless, one can consider a positive exponential potential also for
$\xi \neq 0$, and in this case the necessary condition for cosmic
acceleration is
\be
\frac{\phi}{m_{pl}} > \frac{1}{6\xi} \sqrt{ \frac{p}{\pi}}
\;\;\;\;\;\;\;\;\; \left( 0< \xi \leq \frac{1}{6} \right) \; ,
\ee
\be
\frac{\phi}{m_{pl}} < \frac{1}{6| \xi |} \sqrt{ \frac{p}{\pi}}
\;\;\;\;\;\;\;\;\; \left( \xi <0 \right) \; .
\ee

\section{NMC and the slow-roll approximation to inflation}

\setcounter{equation}{0}

For minimal coupling the equations of inflation  are solved in the
slow-roll approximation \cite{KolbTurner,Liddleetal94}, which amounts to
assuming that the solution is approximately  the de
Sitter space 
\be  \label{dS}
\left( H, \phi \right)=\left( \sqrt{ \frac{\Lambda}{3}}, 0 \right) \; ;
\ee
the slow-roll approximation works
because the solution (\ref{dS}) is an attractor of the dynamical equations
for $\xi=0$, and slow-roll inflation is a quasi-de
Sitter expansion \cite{KolbTurner,Liddlereview,LiddleLyth,Liddleetal94}.

The slow-roll approximation is often used to solve also the equations for
$\xi \neq 0$; however it is unknown whether the de Sitter solution
(\ref{dS}) is still an attractor in this case, and the use of the
slow-roll approximation is unjustified unless this question is
affirmatively answered.. 

We adopt the form (\ref{triangtriang}) and
(\ref{tdoubletilde}) of the field equations, which guarantees
the generality of the solution, the covariant conservation of the
stress-energy  tensor, the weak energy condition, and the constancy of the
gravitational coupling. The equations of motion 
can be written as the
Klein-Gordon equation (\ref{KG}),
the trace of the Einstein equations, and the Hamiltonian constraint,
respectively:
\begin{equation}  \label{4}
R=6\left( \dot{H} + 2H^2 \right)= \kappa \left( \rho
-3P \right) \; ,
\end{equation}
\begin{equation}  \label{5}
3H^2 =\kappa\rho\; ,
\end{equation}
where the energy density and pressure are given by 
\begin{equation}
\rho= \frac{\dot{\phi}^2}{2} +3\xi H^2\phi^2 +6\xi
H\phi \dot{\phi} + V  \label{densityy}
\end{equation}
\begin{equation} 
P= \frac{\dot{\phi}^2}{2} - V -\xi \left( 4H
 \phi \dot{\phi}  +2\dot{\phi}^2 +2\phi \ddot{\phi} \right) -\xi \left( 2
\dot{H} +3 H^2 \right) \phi^2 
 \, .  \label{pressuree}
\end{equation}
The equations of motion can be reduced to a two-dimensional system of
first order equations for the variables $H$ and $\phi$
\cite{Gunzigetal99,FosterBlanco},
\be  \label{reduced1}
 -6\dot{H} \left[ 1 + \xi \left( 6\xi -1 \right) \kappa \phi^2
\right]
+\kappa \left( 6\xi -1 \right) \dot{\phi}^2 -12H^2 
+ 12 \xi \left( 1 - 6\xi
\right) \kappa H^2 \phi^2 + 4 \kappa V  
-6\kappa \xi \phi \frac{dV}{d\phi} = 0 \; ,
\ee
\begin{equation}  \label{reduced2}
-\frac{\kappa}{2}\,\dot{\phi}^2 -6\xi\kappa H\phi\dot{\phi}
+ 3H^2 -3\kappa\xi H^2\phi^2 - \kappa
 V =0 \, ;
\end{equation}
then it is clearly convenient to formulate the problem in terms
of the variables $H$ and $\phi$. One can rewrite the system
(\ref{reduced1}) and (\ref{reduced2}) as two equations that explicitly
give the vector field
$\left( \dot{H}, \dot{\phi} \right)$ of the system. In the language
of dynamical systems, the fixed points of this system are the de Sitter
solutions corresponding to constant Hubble function and scalar
field. 
It is straightforward to check that, for $V=\Lambda /\kappa \geq 0$, the
solutions
\be \label{Sitter}
\left( H, \phi \right) = \left( \pm \sqrt{\frac{\Lambda}{3}}, 0
\right) 
\ee
satisfy Eqs.~(\ref{KG}), (\ref{4}), and (\ref{5}) for {\em arbitrary}
values of $\xi$. The slow-roll formalism is only meaningful when applied
around a {\em stable} de Sitter solution (\ref{Sitter}); otherwise, small
perturbations of the background run away from it and from inflation. Hence
one asks whether the solutions (\ref{Sitter}) are stable or unstable fixed
points; the answer is given by a
local stability analysis. When the potential $V=\Lambda/\kappa $ is left
unchanged
the equations for the perturbations $\delta H $ and $\delta \phi$, defined
by
\be
H=H_0 + \delta H\; , \;\;\;\;\;\;\;\;\;\;\; \phi=\delta \phi \; ,
\ee
yield perturbations that decrease exponentially with time for the
expanding solution (\ref{Sitter}) and therefore stability for any $\xi
\geq 0$; there is instability for $ \xi <0$. The
contracting solution (\ref{Sitter}) is unstable for any value of $\xi$.
It is significant that the sign of the coupling constant $\xi$ affects the
stability of the solution.

However, it is more interesting to consider perturbations of the equations
of motion corresponding to a perturbed potential
\be
V ( \phi ) =\frac{\Lambda}{\kappa} + V_0' \, \delta\phi +
\frac{V_0''}{2}\, \delta
\phi^2 +
\frac{V_0'''}{6} \, \delta\phi^3 + \frac{V_0^{(IV)}}{24} \, \delta\phi^4 +
\ldots \; .
\ee
The cosmological constant is then seen as the zeroth order approximation
of
the potential. The density and pressure perturbations are
given by
\be  \label{densitypert}
\delta \rho = \frac{\delta \dot{\phi}^2}{2} +3\xi H_0^2 \, \delta\phi^2
+6\xi H_0 \, \delta\phi \, \delta\dot{\phi} +\frac{V_0''}{2}
\, \delta\phi^2 + \ldots \; ,
\ee
\be  \label{pressurepert}
\delta P = \frac{\delta \dot{\phi}^2}{2} -4\xi H_0
\, \delta\phi \, \delta\dot{\phi} -2\xi \,
 \delta\dot{\phi}^2-2\xi \, \delta\phi \, \delta \ddot{\phi} -3\xi 
H_0^2 \, \delta \phi^2-\frac{V_0''}{2} \, \delta\phi^2 + 
\ldots \; ,
\ee
where ellipsis denote higher order contributions and the Klein-Gordon
equation implies $\left. V_0' \equiv dV/d\phi \right|_0 =0$.  The 
perturbations satisfy the equations of motion 
\be
\delta \ddot{\phi} +3H_0 \, \delta \dot{\phi} + \left( 12 \xi H_0^2 +V_0''
\right) \, \delta\phi + \ldots = 0 \; ,
\ee
\be
\delta H=\frac{\kappa}{6H_0} \left(  \frac{\delta \dot{\phi}^2}{2}  +
3 \xi H_0^2 \delta \phi^2 +6\xi H_0 \,\delta \phi \, \delta\dot{\phi}
+\frac{V_0''}{2} \delta\phi^2 \right) + \ldots  \; .
\ee
By assuming fundamental solutions of the form
\be
\delta\phi=\epsilon \,  \mbox{e}^{\alpha t}
\ee
one finds the algebraic equation for $\alpha$
\be
\alpha^2 +3H_0\alpha +12 \xi H_0^2 +V_0''=0 \; .
\ee
Let us first analyze the stability of the expanding de Sitter solution
(\ref{Sitter}); the fundamental solutions $\delta \phi_{1,2}$
corresponding
to 
\be
\alpha_{1,2}=\frac{3H_0}{2} \left( -1\pm \sqrt{
1-\frac{16\xi}{3}-\frac{4V_0''}{3\Lambda} } \, \right)
\ee
are exponentially decreasing (or constant) when $
1-16\xi/3-4V_0''/3\Lambda $ is not greater than unity, which corresponds
to {\em stability~} and is achieved for $\xi \geq -V_0''/ ( 4\Lambda)
$. There is {\em instability} when $\xi < -V_0''/ ( 4 \Lambda ) $.

Note that, for $\xi=0$, there is stability for $V_0''>0$ which happens,
e.g., when the
potential has a minimum $\Lambda/\kappa $ at $\phi=0$; a solution starting
at any
value of $\phi$ is attracted towards the minimum (in slow-roll if the
potential is sufficiently flat). If instead $V_0''<0$ and the potential
has a maximum, the solution starting at $\phi =0$ runs away from it.

When $\xi \neq 0$, the potential and the NMC term $\xi R \phi^2/2$
balance; 
if $V( \phi) $ has a minimum $V_0=\Lambda/\kappa $ at $\phi=0$ the
solution is unstable for large negative values of $\xi$. 
If instead the potential has a maximum $\Lambda /\kappa$ at $\phi=0$,
then the expanding de Sitter solution is unstable for any negative $\xi$ 
and stable only for $-4\Lambda \xi \leq V_0'' <0$. Again, the stability
character of
the de Sitter solution is
determined not only by the shape of the potential, but also by the value
of the coupling constant $\xi$. This analysis makes exact the
previous qualitative considerations of
Refs.~\cite{Abbott81,FakirUnruh92ApJ,FutamaseMaeda89,Faraoni96} on the
balance between $\xi R \phi^2/2 $ and $V( \phi)$, and is not limited to
the case in which $V( \phi)$ has an extremum at $\phi=0$.

The contracting de Sitter solution (\ref{Sitter}) is unstable for any
value
of $\xi$, as is deduced by repeating the analysis above.

\section{Conformal transformation techniques}

\setcounter{equation}{0}

Conformal transformation techniques are often used to reduce the study of
a cosmological scenario with a nonminimally coupled scalar field to the
problem of a minimally coupled field, with considerable mathematical
simplification (see Ref.~\cite{FaraoniGunzigNardone99} for a review). The
``Jordan
conformal frame'' in which the scalar field couples nonminimally to the
Ricci curvature is mapped into the ``Einstein frame'' in which the
(transformed) scalar is minimally coupled. The two frames are not
physically equivalent, and care must be taken in applying conformal
techniques \cite{Dick98,FaraoniGunzigNardone99,FaraoniGunzigIJTP}.

The conformal transformation is given by
\be  \label{Emetric}
g_{ab} \longrightarrow \tilde{g}_{ab}=\Omega^2 g_{ab} \; ,
\ee
where
\be  \label{Omega}
\Omega=\sqrt{1-\kappa \xi \phi^2} \; ,
\ee
and the scalar field is redefined according to
\be  \label{Escalar}
d \tilde{\phi}=\frac{\sqrt{1-\kappa \xi ( 1-6\xi ) \phi^2}}{1-\kappa
\xi\phi^2} \, d\phi  \; .
\ee

The ``new'' scalar $\tilde{\phi} $ in the Einstein frame $\left(
\tilde{g}_{ab}, \tilde{\phi} \right) $ is minimally coupled,
\be
\tilde{\Box} \tilde{\phi} - \frac{d \tilde{V}}{d\tilde{\phi}}=0 \; ,
\ee
where 
\be   \label{Vtilde}
\tilde{V} \left( \tilde{\phi} \right) = \frac{ V \left[  \phi \left(
\tilde{\phi} \right) \right]}{\left( 1-\kappa\xi\phi^2 \right)^2}
\ee
and $\phi=\phi \left( \tilde{\phi} \right) $ is obtained by integrating
and inverting
Eq.~(\ref{Escalar}).  The conformal transformation technique  is useful to
solve the equations of cosmology in the Einstein frame and then map  the
solutions $\left( \tilde{g}_{ab}, \tilde{\phi} \right)$  back into the
physical solutions $\left( g_{ab},
\phi \right)$ of the Jordan frame with NMC.  Although from the
mathematical
point of view it
is convenient to obtain exact solutions with NMC in this way (see e.g.
Ref.~\cite{Abreuetal94}), in general the procedure is not very interesting
from the physical
point of view. In fact, one starts from a known solution for a
potential $ \tilde{V} \left( \tilde{\phi} \right) $ motivated by particle
physics in the unphysical Einstein frame, and one obtains a solution
in the physical Jordan frame which corresponds to a potential $V( \phi )$
with no physical justification and, therefore, not very interesting.
Furthermore, if a solution is inflationary in one frame, its
conformally transformed counterpart in the other frame is not necessarily
inflationary.  
To give an example, we consider a conformally coupled scalar field.
Starting with the potential $\tilde{V} \left( \tilde{\phi}
\right)
=\lambda \tilde{\phi}^4$ in the Einstein frame, one integrates
Eq.~(\ref{Escalar}) and uses Eq.~(\ref{Vtilde}) to obtain
\be  \label{newV}
V( \phi )= \left( \frac{3}{2\kappa} \right)^2 \lambda \left(
1-\frac{\kappa}{6} \phi^2 \right)^2 \ln^4 \left[ \frac{
\sqrt{\kappa/6}\, \phi +1 }{\sqrt{\kappa/6}\, \phi -1} \right] \; .
\ee
While the quartic potential in the unphysical Einstein frame  is
everyday's routine, one would be hard put to justify the potential
(\ref{newV}).

There is, however, a meaningful situation in which an inflationary
solution is mapped into another inflationary solution by the conformal
transformation: the slow-roll approximation described in the previous
section. To prove this statement, one begins by noting that an exact de
Sitter solution (\ref{Sitter}) is invariant under the conformal
transformation (\ref{Emetric}), (\ref{Omega}), and (\ref{Escalar}). In
fact, when $\phi$ is constant, Eq.~(\ref{Emetric}) reduces to a rescaling
of the
metric by a constant factor (which can be absorbed into a coordinate
rescaling), and the scalar $\phi$ is mapped into another constant
$\tilde{\phi}$.  Moreover, it is proven in Sec.~6 that a de Sitter
solution is an attractor point in the phase space for
suitable values of the coupling constant $\xi$,  with nonminimal
coupling as well
as with minimal coupling; hence, for these suitable values of $\xi$,  the
conformal transformation maps an attractor of the Jordan frame into an
attractor of the Einstein frame. It is therefore meaningful to consider
the slow-roll approximation to inflation in both frames.

In the Jordan frame the Hubble parameter is given by
\be
a=a_0 \exp \left[ H(t) t \right] 
\; ,
\ee
\be  \label{expansionH}
H(t) =H_0+\delta H (t) \; ,
\ee
where $H_0$ is constant and $\left| \delta H \right| << \left| H_0
\right| $. In the
Einstein frame one has the line element
\be
d\tilde{s}^2 =\Omega^2 ds^2=-d{\tilde{t}}^2+{\tilde{a}}^2 \left( dx^2
+dy^2
+dz^2 \right) \; ,
\ee
where $d\tilde{t}=\Omega dt $ and $\tilde{a} =\Omega a$.
The Hubble parameter in the Einstein frame is 
\be
\tilde{H} \equiv \frac{1}{\tilde{a}} \frac{d\tilde{a}}{d\tilde{t}}
=\frac{1}{\Omega} \left( H+\frac{\dot{\Omega}}{\Omega} \right) 
\; ,
\ee
where an overdot denotes differentiation with respect to the Jordan frame
comoving time $t$. For an exact de Sitter solution $H=$const. implies
$\tilde{H}=$const. and vice-versa. A slow-roll inflationary solution in
the Jordan frame satisfies Eq.~(\ref{expansionH}) and
\be
\phi (t) =\phi_0+\delta \phi (t) \; ,
\ee
where $\phi_0$ is constant and $\left| \delta H \right|<< \left| H_0
\right| $, 
$ \left| \delta \phi \right| << \left| \phi_0 \right| $; the corresponding
Einstein frame quantities are
\be
\tilde{H}= \frac{1}{\sqrt{ 1-\kappa \xi \phi_0^2}} \left(
H_0 +\delta H + \frac{\kappa \xi \phi_0 H_0 }{1-\kappa \xi \phi_0^2} \,
\delta \phi -\frac{\kappa \xi \phi_0 }{1-\kappa \xi \phi_0^2} \, \delta
\dot{\phi} \right) =\tilde{H}_0 + \delta \tilde{H} 
\ee
and
\be
\tilde{\phi}=\tilde{\phi}_0 + \delta \tilde{\phi} 
\ee
where, to first order, 
\be
\tilde{H}_0=\frac{H_0}{\sqrt{ 1-\kappa \xi \phi_0^2}} \; ,
\ee
\be   \label{715}
\frac{\delta\tilde{H}}{\tilde{H}} =\frac{\delta H}{H_0} +
\frac{\kappa \xi \phi_0^2}{1-\kappa\xi\phi_0^2} \left( \frac{\delta
\phi}{\phi_0} -\frac{\delta \dot{\phi}}{H_0 \phi_0} \right) \; ,
\ee
\be
\delta \tilde{\phi}= \frac{\sqrt{ 1-\kappa \xi  \left( 1-6\xi
\right) \phi_0^2}}{1-\kappa\xi \phi_0^2} \, \delta \phi  \; .
\ee

The smallness of the Jordan frame quantities in Eq.~(\ref{715})
guarantees the
smallness of the deviation from a de Sitter solution 
$\delta \tilde{H}/ \tilde{H} $ in the Einstein frame; slow-roll
inflation in the Jordan frame implies slow-roll inflation in the Einstein
frame. The converse is not true, as shown in Ref.~\cite{Abreuetal94} in
the
special case $\xi=1/6$, and therefore some caution must be taken when
mapping back solutions from the Einstein to the Jordan frame. 
These considerations are relevant for the calculation of density and
gravitational 
wave perturbations with NMC aimed at testing NMC inflation with 
present and future satellite observations  
\cite{FakirHabib93,Kaiser95a,Kaiser95b,Hwang90,HwangNoh98}. One must
take special care when computing quantum fluctuations and applying
the conformal transformation (\ref{Emetric}), (\ref{Omega}) and
(\ref{Escalar}) since, in general, the vacum state of one conformal frame
is changed
into a different state in the other frame
\cite{FujiiNishioka90,MakinoSasaki91,FakirHabib93,FabrisTossa97}.

The conformal transformation is only defined for $\xi <0 $ and, if
$\xi
>0$, for values of $\phi$ such that $\phi \neq \pm \phi_c = \pm \left(
\kappa \xi \right)^{-1/2}$. For large
values of $\xi$, this is a serious limitation on the usefulness of
conformal transformation techniques. When $\xi >0$ and the nonminimally
coupled scalar field approaches the
critical values $\pm \phi_c $,
$\tilde{g}_{ab} $ degenerates, $\tilde{\phi}$ diverges and the
conformal transformation technique cannot be applied.  This happens when
$\phi \simeq 0.199 \, 
\xi^{-1/2} m_{pl} $, which induces the unreasonable constraint $\left| 
\phi \right| < 0.2 \, m_{pl} $ if $\xi $ is of order unity (for example,
chaotic inflation requires $\phi$ larger than about $5\,m_{pl}$
\cite{FutamaseMaeda89}). In particular, for strong
positive coupling $\xi >>1$, the critical value $\left| \phi_c \right|$ 
corresponds to very low energies.

The conformal transformation technique cannot provide solutions
with $\phi$ crossing the barriers $
\pm \phi_c$, even when such solutions are physically admissible. In this
sense, the conformal technique has the same limitations of the form of the
field equations (\ref{efe2}) and (\ref{efe3}) discussed in Sec.~2. The
transformed scalar $\tilde{\phi}$ in the Einstein frame can be explicitly
expressed in terms of $\phi$ by integrating Eq.~(\ref{Escalar}),
\be  \label{phif}
\tilde{\phi}= \sqrt{ \frac{3}{2\kappa}} \ln \left[ 
\frac{ \xi \sqrt{6\kappa\phi^2} + \sqrt{1-\xi \left( 1-6\xi \right) \kappa
\phi^2}}
{\xi \sqrt{6\kappa\phi^2} -\sqrt{1-\xi \left( 1-6\xi \right) \kappa
\phi^2}}  \right] + f \left( \phi \right) \; ,
\ee
where 
\be  \label{phi1}
f \left( \phi \right) = 
\left( \frac{1-6\xi}{\kappa \xi} \right)^{1/2} \arcsin
\left( \sqrt{ \xi \left( 1-6\xi \right)  \kappa\phi^2} \right)
\ee
for $0 < \xi \leq 1/6 $ and
\be  \label{phi2}
f \left( \phi \right) = 
\left( \frac{6\xi -1 }{\kappa \xi} \right)^{1/2} 
\mbox{ arcsinh}
\left( \sqrt{ \xi \left( 6\xi -1 \right)  \kappa \phi^2} \right) 
\ee
for $ \xi \geq 1/6 $.
Eqs.~(\ref{phif})-(\ref{phi2}) show that $\tilde{\phi}
\longrightarrow
\pm \infty$ in the Einstein frame as $\phi \longrightarrow \pm \phi_c$ in
the Jordan frame. Any
nonminimally coupled solution $\phi$ crossing the barriers $\pm \phi_c$
cannot be found by applying the conformal transformation
technique.

An explicit example of such a solution is the one corresponding to a
nonminimally  coupled scalar field which is constant and equal
to one of the critical values.  In this case the field equations
(\ref{KG}), (\ref{4}), and (\ref{5}) yield 
$ R=6\left( \dot{H}+2H^2 \right) =0 $
and $V=0$.  The solution
\be
a=a_0 \sqrt{ t-t_0 } \; , \;\;\;\;\;\;\;\;\;\;\; \phi=\pm \,
\frac{1}{\sqrt{\kappa\xi}} \;\;\;\;\;\;\;\; \left( \xi >0 \right) \; ,
\ee
corresponds to the vanishing of the trace of the energy-momentum tensor
$\tilde{T}_{ab} \left[ \phi \right] =\tilde{T}_{ab}^{(total)} $ in
Eqs.~(\ref{triangtriang}) and (\ref{tdoubletilde}), and to the radiation
equation of state $P=\rho/3$.

\section{Discussion and conclusions}

\setcounter{equation}{0}

Scalar fields are a basic ingredient of particle physics and cosmology and
many arguments strongly suggest that a scalar field must couple
nonminimally to the Ricci curvature of spacetime in the theories of
gravity and of the scalar field used to build most scenarios of inflation
and quintessence. Therefore, one cannot ignore NMC in these models. 
In this paper we approach several topics in the physics of NMC, 
from a general (i.e. not limited to a specific
potential $V( \phi )$) point of view.

First, it is shown that the possible forms of writing the field equations
are not equivalent, and it is pointed out that some of them lead to loss
of generality and to a restricted class of solutions. This is not a
problem when one focuses on a specific solution $\left( g_{ab}, \phi 
\right) $, but it compromises studies that aim at generality like, e.g.,
the dynamical system analysis of the equations of cosmology in the phase
space. Further, a shadow is cast on the reality of the time-variability of
the effective gravitational constant $G_{eff}(t)$ in many cosmological
scenarios using NMC. In fact, the time-variability may be removed by
passing to a different form of the field equations, and this
interpretation
problem deserves attention in the future.

The conservation equations for the different forms of the field equations
are discussed, and it emerges that different formulations lead to
different definitions of the energy density and pressure of the scalar
field. As a result of this fact, the problem of whether a
nonminimally coupled
scalar field satisfies the weak energy condition becomes fuzzy. In this
paper, the form
of the field equations (\ref{triangtriang}) and (\ref{tdoubletilde}) is
preferred because {\em i})~it does not lead to loss of generality, {\em
ii})~the stress-energy tensor (\ref{tdoubletilde}) is covariantly
conserved and satisfies the weak energy condition in spatially flat and
closed FLRW universes, and {\em iii})~the
gravitational coupling is constant, and there are no interpretation
problems with a time-varying $G_{eff}(t)$.

The crucial feature of inflationary and quintessence models, i.e. the
acceleration of the universe, is studied when the universe is dominated
by a nonminimally coupled scalar field. The inclusion of the
NMC term $\xi R \phi^2 /2 $ in the Lagrangian density 
seems to make it harder to achieve cosmic acceleration for
most potentials that are known to be inflationary when $ \xi =0$.
This conclusion derives from the dynamical equations for the scalar field
and the scale factor and does not rely upon the
slow-roll approximation, nor does it arise from independent consistency
requirements of the kind discussed in Ref. \cite{Faraoni96}. In addition
to the dynamical arguments, one must keep in
mind that a given inflationary scenario must be consistent
with the theoretical prescriptions for the value of $ \xi $, which further
constrain the known scenarios \cite{Faraoni96,Calgary}.
Fine-tuning arguments or, in other words, the genericity of inflation,
are also an issue \cite{FutamaseMaeda89,ALO90}.

The NMC term $\xi R \phi^2 /2 $ can balance a suitable scalar field
potential $V( \phi)$ and induce cosmic acceleration 
with a wider class of potentials than it is normally considered. However,
the NMC term in the Lagrangian cannot completely 
substitute for a potential and induce an acceleration epoch 
when $V=0$. We have proved this statement in Sec.~4 for values of the
coupling constant  $\xi \leq 1/6$, but we could not reach a conclusion
for $\xi > 1/6$. A new solution for $\xi=0$ expanding from a big-bang
singularity and quickly approaching a de Sitter space is also
presented.

Since almost all the inflationary scenarios proposed to date are based on
the slow-roll approximation, the role of de Sitter solutions (which are
fixed points) as attractor points in the phase space is crucial. We study
this issue in the presence of NMC and we find that the stability of the
expanding de Sitter solution (\ref{Sitter}) is determined not only by the
shape of the potential $V( \phi )$ (as is the case of minimal coupling),
but also by the value of the coupling constant $\xi$. A more general
analysis including perturbations which are space-dependent or 
anisotropic is needed to confirm the stability; however, our local
perturbation analysis is sufficient to establish instability for 
\be
\xi < \xi_0 \equiv -\, \frac{V''_0 }{4\Lambda} \; ,
\ee
where $\Lambda =\kappa V_0 $.
Note that $\xi_0=-\eta_0 /4$, where $\eta = V''/(
\kappa V )$ is one of the slow-roll parameters used in the slow-roll
approximation for minimal coupling \cite{LiddleLyth}.

Contracting de Sitter solutions are always unstable, as in the $\xi=0$
case.  These considerations set precise limits on the domain
in which the slow-roll
approximation is meaningful in the presence of NMC, and is fundamental for
the computation of scalar and tensor perturbations. Ultimately, the
amplitudes and spectral indices of these perturbations are the
predictions of the theory to be compared with observations of the cosmic
microwave background. 

Conformal transformation techniques are widely used in scalar field
cosmology \cite{FaraoniGunzigNardone99} and it is useful to
clarify their link with the slow-roll approximation. We prove that
slow-roll inflation in the physical Jordan frame (in which the scalar
field is nonminimally coupled) implies slow-roll inflation in the
unphysical Einstein frame (but not vice-versa), and make explicit the
limitations intrinsic to the use of conformal transformations. Analytic
examples are given in which the conformal transformation method cannot be
applied.

Recently, there has been a great deal of work on NMC in both inflationary
and quintessence models; this paper justifies certain assumptions and
methods used, solves some of the problems posed, and provide
caveats on difficulties that were overlooked. Our considerations will be
applied in the future to specific models; other areas in which NMC is
relevant include quantum cosmology, classical and quantum wormholes,
and the
stability of boson stars.

\section*{Acknowledgments}

It is a pleasure to thank S. Odintsov and E. Gunzig for useful 
discussions, and a referee for valuable contributions to Sec.~4.3. This
work was supported by the EEC grant PSS*~0992 and by OLAM, Fondation pour
la Recherche Fondamentale, Brussels.


\begin{thebibliography}{99}
\bibitem{footnote1} Nevertheless, even
in $R^2$ inflation, a scalar field is sometimes added to the
scenario in order to increase the amount of inflation \cite{Maeda89}.
\bibitem{footnote2} The metric signature 
is --~+~+~+ and $G$ denotes Newton's constant. The speed of light and
Planck's constant assume the value
unity and $m_{pl}=G^{-1/2}$ is the Planck mass.  The components of the
Ricci tensor are given in terms of the
Christoffel symbols
$\Gamma_{\alpha\beta}^{\delta}$ 
by $R_{\mu\rho}=
\Gamma^{\nu}_{\mu\rho ,\nu}-\Gamma^{\nu}_{\nu\rho ,\mu}+
\Gamma^{\alpha}_{\mu\rho}\Gamma^{\nu}_{\alpha\nu}-
\Gamma^{\alpha}_{\nu\rho}\Gamma^{\nu}_{\alpha\mu} $, 
$\Box \equiv g^{ab}\nabla_{a}\nabla_{b}$, and the abstract index 
notation \cite{Wald} is adopted.
\bibitem{footnote3} The value $1/6$
of the coupling constant $\xi$ makes the Klein-Gordon
equation (\ref{KG}) in four spacetime dimensions conformally invariant if
$V=0$ or $V=\lambda \phi^4$ \cite{Wald,BirrellDavies80}.
\bibitem{footnote5} For de Sitter
solutions, the Klein-Gordon equation is not independent of the Einstein
field equations \cite{EllisMadsen,Gunzigetal99}.
\end{thebibliography}
\end{document}